\documentclass[8.5pt,twoside,twocolumn]{article}
\usepackage[super,sort&compress,comma]{natbib} 
\usepackage{mhchem}
\usepackage{times,mathptmx}
\usepackage{times}
\usepackage{sectsty}
\usepackage{balance}

\usepackage{graphicx}
\usepackage{lastpage}
\usepackage[format=plain,justification=raggedright,singlelinecheck=false,font=small,labelfont=bf,labelsep=space]{caption} 
\usepackage{fancyhdr}
\usepackage{lipsum}
\begin{document}

\thispagestyle{plain}
\fancypagestyle{plain}{
\renewcommand{\headrulewidth}{1pt}}
\renewcommand{\thefootnote}{\fnsymbol{footnote}}
\renewcommand\footnoterule{\vspace*{1pt}
\hrule width 3.4in height 0.4pt \vspace*{5pt}} 
\setcounter{secnumdepth}{5}

\makeatletter 
\def\subsubsection{\@startsection{subsubsection}{3}{10pt}{-1.25ex plus -1ex minus -.1ex}{0ex plus 0ex}{\normalsize\bf}} 
\def\paragraph{\@startsection{paragraph}{4}{10pt}{-1.25ex plus -1ex minus -.1ex}{0ex plus 0ex}{\normalsize\textit}} 
\renewcommand\@biblabel[1]{#1}            
\renewcommand\@makefntext[1]%
{\noindent\makebox[0pt][r]{\@thefnmark\,}#1}
\makeatother 
\renewcommand{\figurename}{\small{Fig.}~}
\sectionfont{\large}
\subsectionfont{\normalsize} 

\fancyfoot{}
\fancyfoot[RO]{\footnotesize{\sffamily{1--\pageref{LastPage} ~\textbar  \hspace{2pt}\thepage}}}
\fancyfoot[LE]{\footnotesize{\sffamily{\thepage~\textbar\hspace{3.45cm} 1--\pageref{LastPage}}}}
\fancyhead{}
\renewcommand{\headrulewidth}{1pt} 
\renewcommand{\footrulewidth}{1pt}
\setlength{\arrayrulewidth}{1pt}
\setlength{\columnsep}{6.5mm}
\setlength\bibsep{1pt}

\twocolumn[
  \begin{@twocolumnfalse}
\noindent\LARGE{\textbf{Cell detachment and label-free cell sorting using modulated surface acoustic waves (SAW) in droplet-based microfluidics$^\dag$}}
\vspace{0.6cm}

\noindent\large{\textbf{Adrien Bussonni\`ere,\textit{$^{a}$} Yannick Miron,\textit{$^{b}$} Micha\"el Baudoin,$^{\ast}$\textit{$^{a}$} Olivier Bou-Matar,\textit{$^{a}$} Michel Grandbois,\textit{$^{b}$} Paul Charette$^{\ast}$\textit{$^{c}$} and Alan Renaudin$^{\ast}$\textit{$^{c}$}}}\vspace{0.5cm}



\vspace{0.6cm}

\noindent \normalsize{We present a droplet-based surface acoustic wave (SAW) system designed to viably detach biological cells from a surface and sort cell types based on differences in adhesion strength (adhesion contrast), without the need to label cells with molecular markers. The system uses modulated SAW to generate pulsatile flows in the droplets and efficiently detach the cells, thereby minimizing SAW excitation power and exposure time. As a proof-of-principle, the system is shown to efficiently sort HEK 293 from A7r5 cells based on adhesion contrast. Results are obtained in minutes with sorting purity and efficiency reaching 97 \% and 95 \%, respectively.} 
\vspace{0.5cm}
 \end{@twocolumnfalse}
  ]
\footnotetext{\textit{$^{a}$~LIA LEMAC/LICS, Institut d'El\'ectronique de Micro\'electronique et de Nanotechnologies (IEMN) UMR CNRS 8520, Universit\'e Lille 1 and EC Lille, Avenue Poincar\'e, BP 60069, 59652 Villeneuve d'Ascq, France; E-mail: michael.baudoin@univ-lille1.fr}}
\footnotetext{\textit{$^{b}$~D\'epartement de Pharmacologie, Falcult\'e de m\'edecine et des sciences de la sant\'e, Universit\'e de Sherbrooke, 3001, 12e Avenue Nord Sherbrooke, Sherbrooke, Qu\'ebec J1H 5N4, Canada.}}
\footnotetext{\textit{$^{c}$~UMI LN2 3463 CNRS, Biophotonics and Optoelectronics laboratory, 3IT, Universit\'e de Sherbrooke, Sherbrooke, Qu\'ebec J1K 0A5, Canada. Fax: +1 (819) 821-7937; Tel: +1 (819) 821-8000 x65788;  E-mail: Alan.Renaudin@USherbrooke.ca or Paul.G.Charette@usherbrooke.ca}} 

\section{Introduction}
Cell sorting is critical for many biological and biomedical applications such as cell biology, biomedical engineering, diagnostics and therapeutics. Indeed, numerous biological analyses are based on the separation of different cell types harvested from a raw heterogeneous sample such as whole blood. Fluorescence-activated cell-sorting (FACS)\cite{Herzenberg1976} and magnetic-activated cell-sorting (MACS)\cite{Miltenyi1990} are well established methods for cell and particle sorting known for their high-throughput and specificity. Both methods, however, require pre-processing for cell tagging with markers, an important time and cost expense for some applications.

In contrast, by making use of differences in the intrinsic physical properties of cells (size, density, adhesion strength, stiffness, electrical and optical polarizability), label-free sorting methods do not require molecular tagging. Compared to FACS and MACS, however, the specificity of label-free methods is often limited owing to insufficient contrast in physical properties, thereby restricting their widespread use. As with tagging-based systems, label-free cell-sorting methods have been implemented in microfluidic devices\cite{Gossett2010} using techniques such as deterministic lateral displacement\cite{Huang2004}, hydrodynamic filtration\cite{Yamada2005}, dielectrophoresis (DEP)\cite{Becker1995,Vahey2008}, optical lattices\cite{Macdonald2003,Jonas2008}, stiffness separation\cite{Carlson1997,Wang2013a}, acoustophoresis\cite{Laurell2007,Ai2013} and adhesion-based sorting\cite{Sin2005,Kwon2007,Didar2010}. In some cases, such as for sorting based on adhesion to the substrate, performance has been improved with surface nanostructuring\cite{Kwon2007}  and bio-functionalization\cite{Didar2010} to enhance adhesion contrast between cell types. 

In addition to sorting, the dynamics of cell detachment from solid surfaces is of interest in and of itself, either to harvest cells or to study the mechanisms of cell adhesion to surfaces. Cell dissociation and detachment from a solid substrate is normally achieved by cleaving bonding proteins with trypsin\cite{Kuhne1877}. This process is quite aggressive as cells can be damaged if left exposed to trypsin for too long and a post-treatment rinsing step is required. In contrast, cell detachment based on microfluidic effects alone requires no external agents or rising. Cell detachment under constant fluid shear stresses has been demonstrated using spinning discs\cite{Weiss1961}, flow chambers\cite{Usami1993} and more recently using surface acoustic wave (SAW)-actuated flow\cite{Schneider2008,Hartmann2013a}. 

The miniaturization of cell-manipulation methods has led to their integration into lab-on-chip (LOC) platforms, where cell detachment and sorting have been widely investigated in flow-based microchannel formats\cite{Usami1993,Lu2004,Lenshof2010}. Comparatively few studies\cite{Fan2008,Shah2009,Shah2010,Baret2009}, however, have explored cell separation or detachment in droplet-based microfluidics as in \textit{digital microfluidics} (DMF)\cite{Berthier2013}. Indeed, the physics of microfluidics in droplets are completely distinct from flow-through closed-channels systems. Microfluidics properties such as bulk and surface modes of vibration, which are unique to droplet-based systems, can be exploited to great effect. In general, unlike continuous flow systems which are often optimized for high-volume cell-sorting, droplet-based systems are best suited to studies of cell properties in small populations of cells, such as cell adhesion modulation mechanisms which are highly complex and of wide-ranging interest\cite{Parsons2010,Safran2005}.

Droplet actuation in DMF is generally accomplished with surface acoustic waves (SAW)\cite{Wixforth2004,Friend2011,Yeo2014}, electrowetting on dielectrics (EWOD)\cite{Nelson2012} or dielectrophoresis (DEP)\cite{Pollack2000}. SAW-based DMF have been used in a range of biological applications\cite{Guttenberg2005,Lyford2012,Renaudin2010} to implement functions as diverse as mixing\cite{Alghane2012}, droplet displacement\cite{Brunet2010,Renaudin2006}, atomization\cite{Qi2008}, as well as for particle and cell manipulation\cite{Li2007}. Though EWOD has been used successfully to manipulate cells\cite{Shah2009}, large or strongly-adhering cells are difficult to detach and/or transport with the relatively weak electrowetting forces exerted by EWOD.  In the case of DEP, the oscillatory forces applied to the cell are subject to changes in the physical composition of the membrane\cite{Gagnon2008}, which may or may not be desirable depending on the nature of the experiment. The strong electrical fields involved in DEP also have the potential to alter cell membrane characteristics\cite{Sengupta2009}.

Recently\cite{Bussonniere}, we presented preliminary results on the use of SAW-based fluid actuation in droplets to detach biological cells from a surface. We showed that under continuous SAW excitation, fully confluent cell layers could be detached \textit{en masse} at sufficiently high SAW power, whereas isolated cells were very resistant to detachment, even at power levels above the threshold for cell viability\cite{Li2009}. In a previous publication, we showed that the acoustic power required to move or deform droplets with SAW can be significantly reduced by using modulated rather than continuous excitation\cite{Baudoin2012}. Based on this work, we demonstrate here that modulated SAW can be used to viably detach cells from a surface and sort cells based on adhesion contrast, without the need for labeling. The experimental results presented show that two distinct cell types can be separated with a final purity of up to 97 \% and an efficiency greater than 95 \%. Results are achieved with characteristic processing times on the order of one minute without adversely affecting cell viability or requiring the cell layer to be fully confluent.

\section{Methods and Materials}
\subsection{Apparatus setup and experimental procedure}
Experiments were run on cells adhered to the surface of a LiNbO$_3$ substrate and immersed in 20 $\mu$l droplets of phosphate buffered saline solution (PBS). 

Rayleigh-type SAW were generated at the surface of the LiNbO$_3$ substrate by applying a 17.1 MHz sinusoidal radio frequency (RF) excitation to interdigitated transducer (IDT) thin-film metal electrodes. The excitation frequency was selected to maximize the energy transmission from the substrate surface to the liquid. The IDT were designed as "electrode width controlled single-phase unidirectional transducers" (EWC-SPUDT)\cite{Hartmann1989}, a configuration that ensures that the acoustic energy is directed solely in the forward direction. The excitation signal was supplied by an RF generator (Agilent, model N9310a) and amplified to 30 dBm using an RF amplifier (Empower, model BBM0D3FEL).

Once transmitted to the fluid, the acoustic waves induce interface stresses and internal flow resulting in deformations of the droplet free surface due to two types of nonlinear effects: acoustic radiation pressure and acoustic streaming\cite{Brunet2010}. As explained below, cyclic droplet deformations were induced by switching the SAW excitation on and off with an appropriate period and duty cycle, resulting in large shear stresses in the fluid causing cells to detach.

\begin{figure}[!h]
\centering
  \includegraphics[height=5cm]{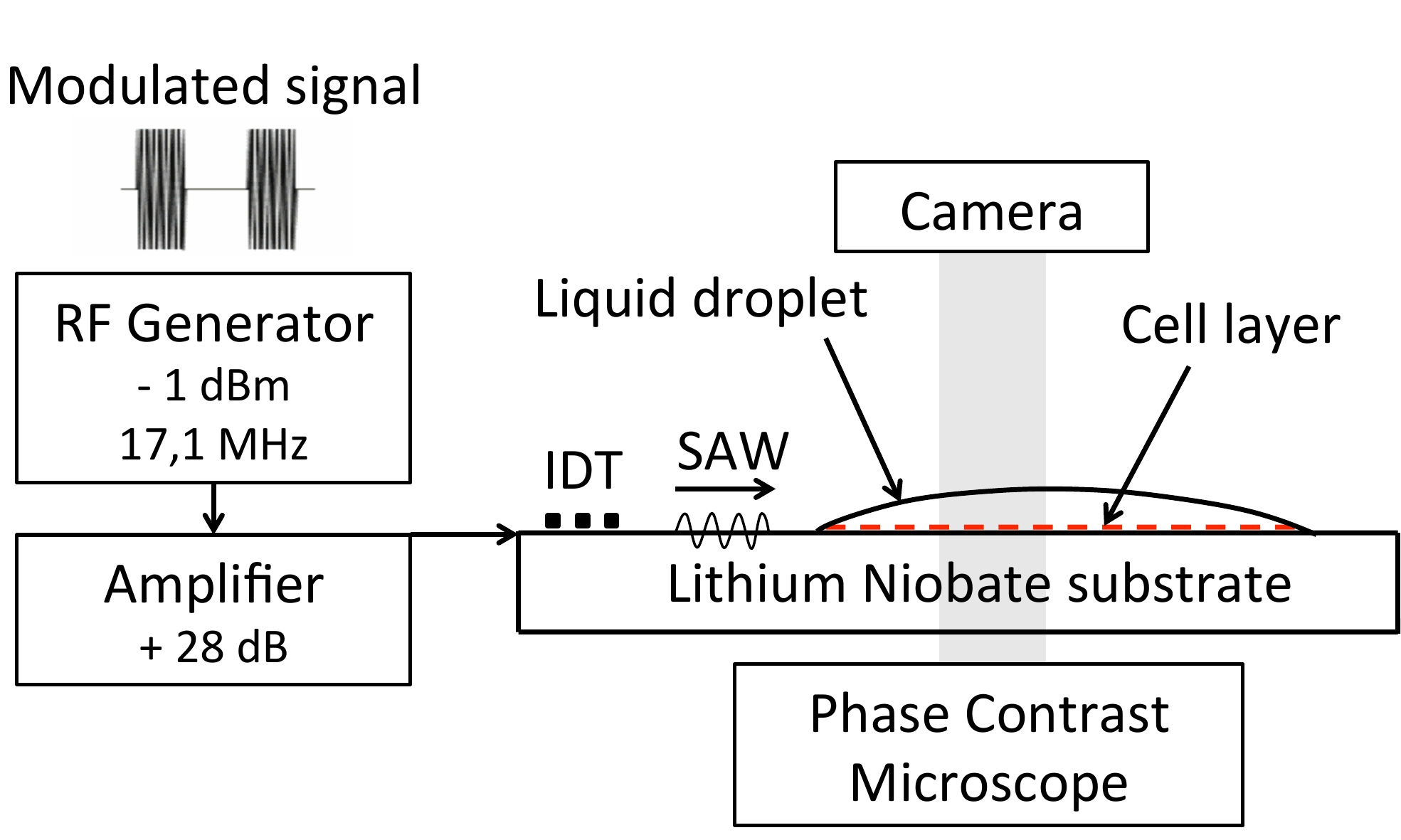}
  \caption{Schematic diagram of the setup showing the cell layer (red dashed line) immersed in a 20 $\mu$l PBS droplet atop a LiNbO$_3$ piezoelectric substrate. The diagram also shows the interdigitated transducers (IDT), RF amplifier and signal generator used for Rayleigh type SAW generation. Cell visualization is achieved using a phase contrast phase microscope, 10$\times$ objective, and CMOS camera.}
  \label{fgr:setup}
\end{figure}

An 8$\times$8 mm$^2$ "cell-attachment zone" was defined by markers patterned onto the LiNbO$_3$ substrates. The devices were mounted under a phase contrast microscope (Motic, model AE 30/31) with a 10$\times$ objective to focus on the cell layer (2 mm diameter field of view). A high-speed 10 bit CMOS camera (PCO, model pco.1200hs) was used to capture video of cell detachment. The setup is depicted in Figure \ref{fgr:setup}.

Images sequences recorded by the camera were processed using ImageJ (NIH, rsbweb.nih.gov/ij) and cell populations were counted using the ImageJ cell counting tool. The two cell types used in the experiments could be easily distinguished in the images based on differences in their morphological characteristics. Following each experiment, cell short term viability assays were performed with Trypan blue exclusion.

\begin{figure}[!h]
\centering
  \includegraphics[height=3.7cm]{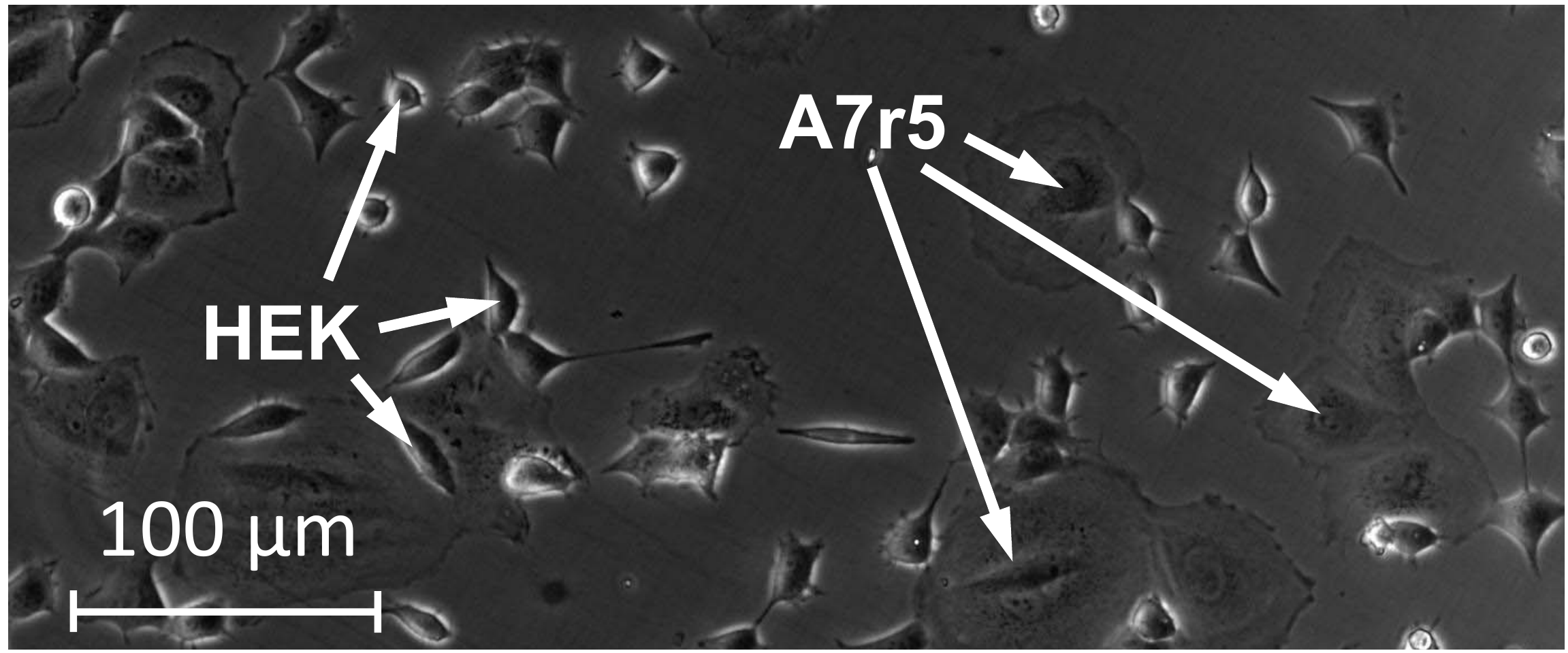}
  \caption{Photograph of vascular smooth muscle cells (A7r5) and human embryonic kidney cells (HEK 293) adhere to the surface of a LiNbO$_3$ substrate after incubation in a re-suspended cell solution at 37$^{\circ}$C. The arrows indicate typical specimens of HEK 293 and A7r5 cells which exhibit significantly different morphological features.}
  \label{fgr:cell}
\end{figure}

\subsection{SAW Device fabrication}
The SAW devices were fabricated from 1 mm thickness X-cut Z-propagating LiNbO$_3$ wafers (Newlight Photonics). This particular cut was chosen for its efficient electromechanical coupling along both Y and Z perpendicular propagation directions (K$_z$ = 4.9 \%, K$_y$ = 3.1 \%), allowing for two-dimensional droplet actuation if required. Indeed, the K$^2$ values along the two directions are greater than for a typical 128$^{\circ}$ Y-X cut LiNbO$_3$ crystal (K$_x$ = 5.5 \%, K$_y$ = 1.2 \%).

The metal IDT electrodes and cell-attachment zone markers were fabricated by photoresist (Shipley S18-13, Microchem) spin-coating on the LiNbO$_3$ wafers, patterning with standard photolithography processes, metal deposition (Ti/Au, 20/200nm), and lift-off.

\subsection{Cells culture and surface preparation}
Two cell lines were used in the experiments: adherent vascular smooth muscle cells (A7r5) and human embryonic kidney cells (HEK 293). These particular cell lines were selected because they have been shown to exhibit surface adhesion strengths comparable to cancerous\cite{Plouffe2008}  and other normal\cite{Palmer2008} cell types. Cell adhesion strength, however, is highly dependent on surface preparation specifics. We chose to adhere cells directly to a bare lithium niobate (LiNbO$_3$), the piezoelectric material most commonly used to generate surface acoustic waves, in order to provide a recognizable point of reference. The adhesion of cells to bare LiNbO$_3$ has been studied by other groups\cite{Rodaite-Riseviciene2013}. As shown below, the fluid shear stresses required to detach cells from a bare LiNbO$_3$ surface are in the same range as that reported by others for surface preparations commonly used for cell studies. If required, LiNbO$_3$ can be readily functionalized for specific surface preparations\cite{Seeger1992,Bennes2008}. Cells were adhered to the device surfaces either by growing the cells directly on the LiNbO$_3$ substrates in the case of single cell line reference experiments, or by incubating the LiNbO$_3$ substrates in a solution of pre-grown re-suspended cells in the case of single and dual cell line experiments. 

A7r5 and HEK 293 cells were grown separately by seeding in 60 mm petri dishes and cultured in growth media (DMEM supplemented with 10\% heat-inactivated fetal bovine serum, 2 mM L-glutamine, 50 IU/mL penicillin, 50$\mu$g/mL streptomycin, Wisent) under an atmosphere of 5\% CO2 at 37$^{\circ}$C for 24h. For experiments requiring cells to be grown directly onto LiNbO$_3$ surfaces, the LiNbO$_3$ substrates were placed at the bottom of the petri dish for the duration of the incubation time.

For cells not grown directly onto the LiNbO$_3$ substrates, single cell line re-suspended cell solutions were prepared by rising the cell cultures in PBS, incubating for 5 min in 500 $\mu$l trypsin-EDTA solution at 37$^{\circ}$C, and re-suspending in 2 ml growth medium to stop the trypsin digestion. After 5 min centrifugation, cells were re-suspended in HEPES-buffered salt solution (HBSS) (20 mM Hepes at pH 7.4, 120 mM NaCl, 5.3 mM KCl, 0.8 mM MgSO4, 1.8 mM CaCl2, and 11.1 mM dextrose). Dual cell line re-suspended solutions were prepared by combining HEK 293 and A7r5 re-suspended solutions in experimentally-determined proportions as required for roughly equal numbers of both cell types to adhere to the surface (Fig. \ref{fgr:cell}). LiNbO$_3$ substrates were placed in a petri dish immersed in re-suspended cell solutions for incubation times ranging from 15 to 90 min to study various adhesion states.
	
Following cell adhesion by either of the two methods above, the LiNbO$_3$ substrates were rinsed in PBS, cells outside the attachment zone were dried and mechanically removed, leaving a uniformly distributed 8$\times$8 mm$^2$ adherent cell layer covered by a thin PBS film. A 20 $\mu$l droplet of PBS was added atop the attachment zone using a calibrated micropipette prior to experiments. Note that due to adherent cell secretions, the cell-attachment zone was highly hydrophilic and the droplets spread across the squared-shaped area with a high contact angle and flat profile.

\begin{figure}[!h]
\centering
  \includegraphics[height=6cm]{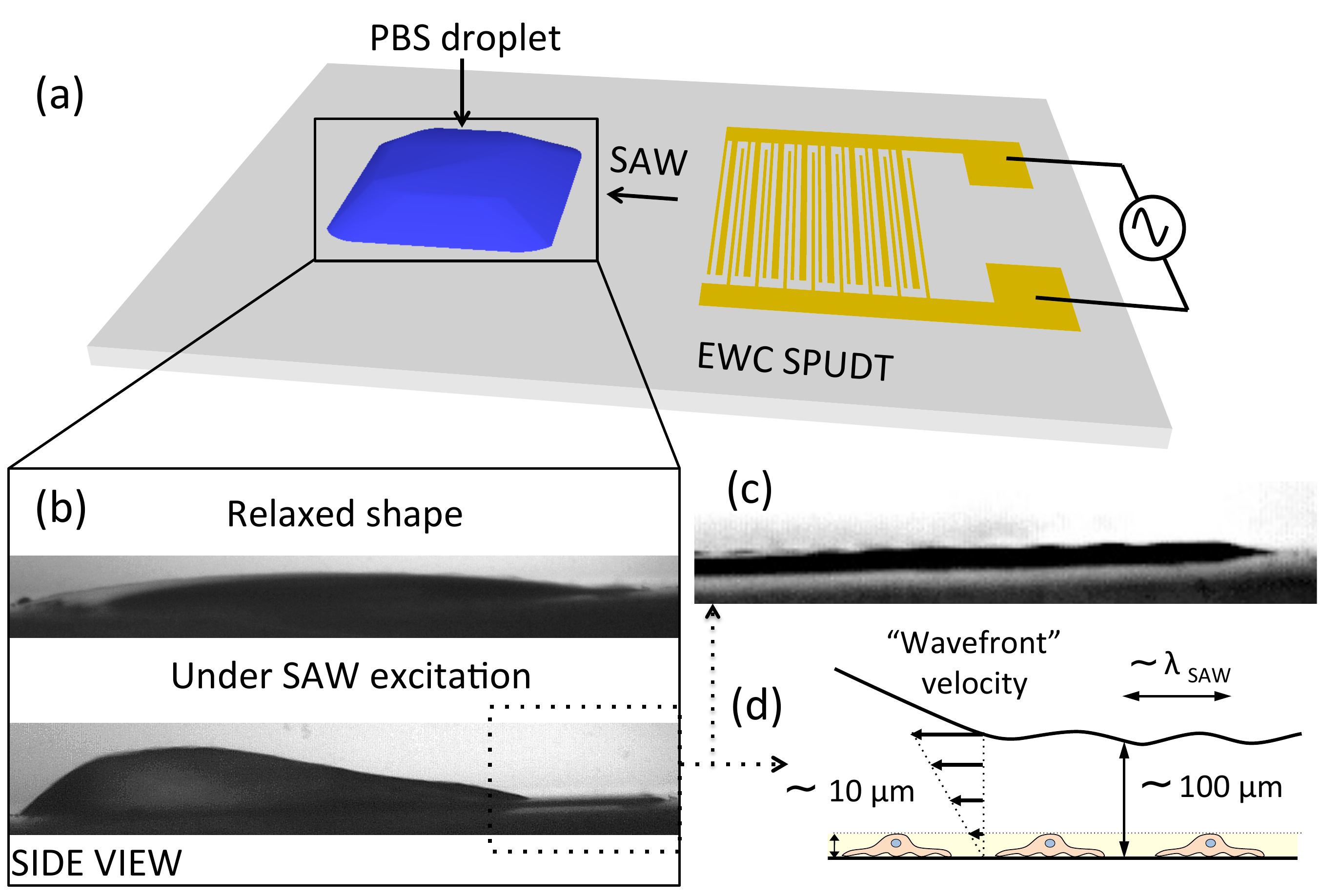}
  \caption{(a) Schematic diagram of the fluid droplet in its relaxed position spread across the 8$\times$8 mm$^2$ droplet positioning zone atop the LiNbO$_3$ substrate, in relation to the SAW electrodes (EWC-SPUDT); (b) side-view camera images showing the two shape extrema of the droplet during one SAW excitation modulation cycle (ON: 50 ms, OFF: 150 ms). (c) Zoom on the tail section of the drop showing a standing wave pattern. (d) Flow profile in the tail section and characteristic lengths.}
  \label{fgr:physics}
\end{figure}

\section{Results and discussion}
\subsection{Droplet dynamics and cell-fluid interaction}
As stated earlier, the purpose of this work was to investigate the potential of modulated SAW fluid actuation to selectively detach and sort cells in a droplet. Fig.\ref{fgr:physics}a shows a schematic diagram of PBS the fluid droplet atop the  8$\times$8 mm$^2$ droplet positioning zone on the LiNbO$_3$ substrate, in relation to the SAW electrodes (EWC-SPUDT). Cyclic deformations of the droplets between a relaxed state and a deformed state (side-view camera images shown in Fig.\ref{fgr:physics}b-top and Fig.\ref{fgr:physics}b-bottom, respectively) were induced by switching the SAW signal on (excitation) and off (relaxation) with an appropriate period and duty cycle. During excitation, internal flow and surface deformation are induced by nonlinear acoustic forces. During relaxation, the potential energy stored as capillary surface energy produces a restoring flow in the opposite direction. Optimal values for the period (200 ms), duty cycle (25 \%), and power (30 dBm) were based on the droplet intrinsic relaxation time, measured experimentally.

\begin{figure}[!h]
  \centering
  \includegraphics[height=8cm]{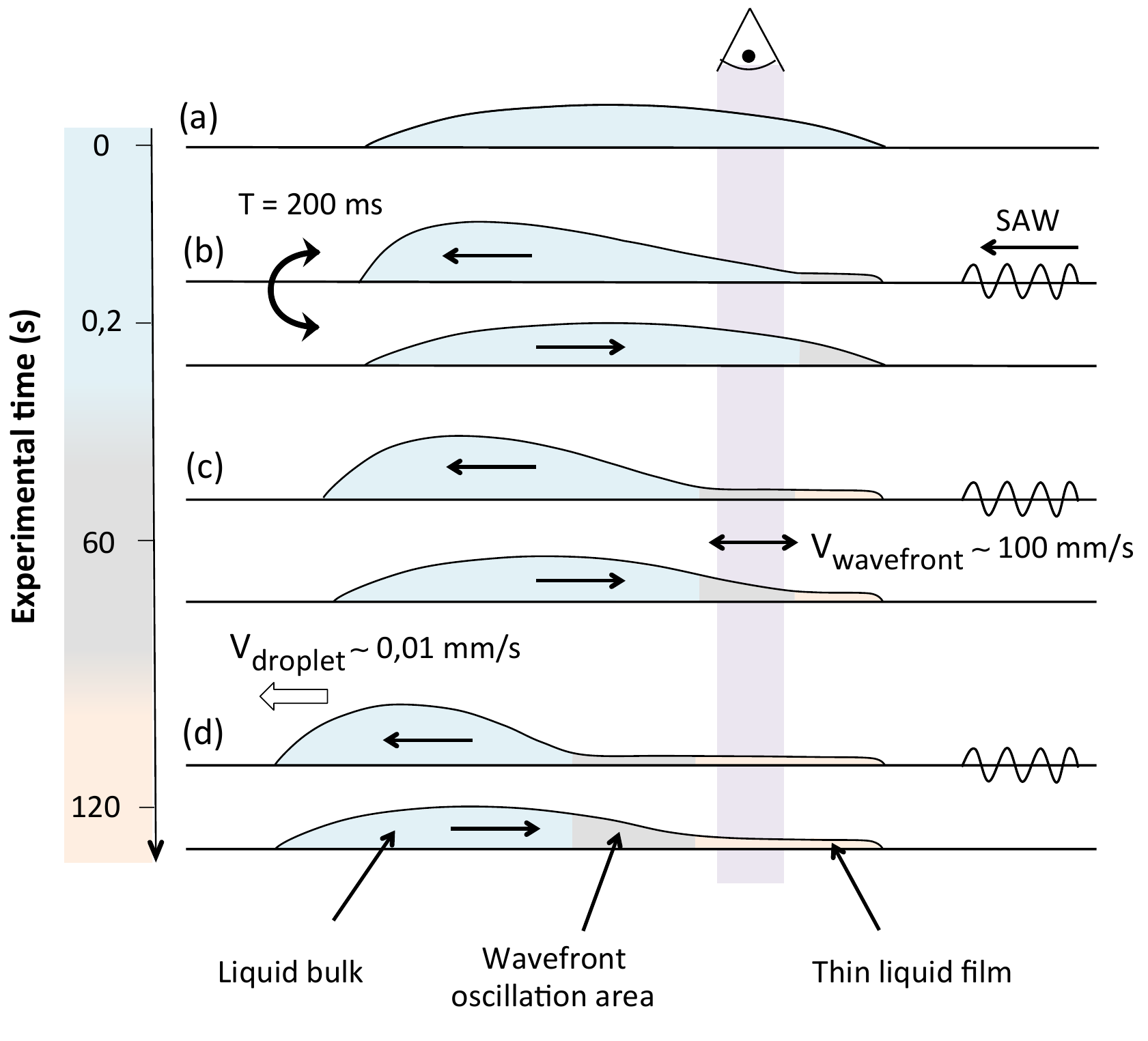}
  \caption{Schematic diagrams of the different droplet fluid dynamics regimes under modulated SAW actuation (ON: 50 ms, OFF: 150 ms). The vertical bar shows the location of the circular area (2 mm diameter) imaged by the microscope; (a) Relaxed shape of the droplet mainly determined by its wetting properties on the cell-covered LiNbO$_3$ substrate; (b) Initial shape extrema pair in each excitation/relaxation cycle. (c) After a sufficient number of cycles, a 150 ms relaxation interval between SAW excitation bursts is no longer sufficient to return the droplets to a symmetric relaxed shape of (b). As a result, three distinct fluid dynamics zones can be distinguished: bulk (blue), transient (gray), and thin-film "tail" (orange); (d) Eventually, as the droplet spreads sufficiently, the field of view of the camera is confined to the thin-film tail.}
  \label{fgr:explication}
\end{figure}

The 50 ms SAW excitation bursts displaced the distal liquid-solid contact line (droplet edge furthest from the SAW electrodes) at a rate of a few microns per cycle while the proximal contact line (droplet edge closest to the SAW electrodes) remained pinned due to the high contact angle hysteresis. As a result, the droplets spread in the direction of SAW propagation at a velocity of $\sim$ 0.01 mm/s. As shown in Fig.\ref{fgr:explication}, this expansion elicited three distinct fluidic regimes in the droplets. Since the camera field of view (vertical column in Fig.\ref{fgr:explication}) was fixed relative to the spreading droplet, the three fluidic regimes swept sequentially across the field of view and thus could be separated in time in the image sequences. 

At the start of the experiments, the expansion was sufficiently small so that the droplets essentially oscillated between the two shape extrema shown schematically in Fig.\ref{fgr:explication}b and in photographs of (Fig.\ref{fgr:physics}b). During SAW excitation (Fig.\ref{fgr:explication}b-top), most of the fluid was displaced in a "bulk" zone at the distal end of the droplet whilst leaving behind a thin film "tail", as previously observed by Rezk \textit{et al} and Collins \textit{et al}\cite{Rezk2012a,Collins2012}. During relaxation, the droplet returned to its symmetric relaxed shape due to surface tension effects (Fig.\ref{fgr:explication}b-bottom).

After a sufficient number of cycles, the distal contact line displacement became significant enough ($\sim$ 0.5 mm) that a 150 ms relaxation interval between SAW excitation bursts was no longer sufficient to return the droplets to a symmetric relaxed shape. As a result, \textit{three} distinct fluid dynamics regimes (Fig\ref{fgr:explication}.c) could be distinguished in the droplets. As before, the extrema were: (1) the \textit{bulk} (shaded blue area), where shear stresses result from a combination of large scale vortices induced by Eckart streaming\cite{Eckart1948} and small scale vortices due to Rayleigh and Schlichting streaming\cite{Rayleigh1884,Schlichting1968} near the viscous boundary layer, and (2) the thin film \textit{tail} (shaded orange area), where Schlichting and Rayleigh streaming generated by a standing wave (as evidenced by periodic patterns in Fig.\ref{fgr:physics}c) are dominant\cite{Rezk2012a}. In between the two, a \textit{transient} zone appeared (gray shaded area), swept by a strong oscillating "wavefront" resulting from the fluid transitions to and from the extrema. 

As shown in the experimental results below, the rate of cell detachment from the surface was consistently highest in the transient zone. Eventually, as the droplets spread sufficiently, cells in the field of view of the camera were confined to the thin film tail (Fig.\ref{fgr:explication}d, shaded orange area).

\subsection{Magnitude of the shear stresses and unsteady forces in the transient zone}
Shear stresses in the fluid acting on the cells were estimated from the image sequences in the transient zone. Assuming no-slip boundary condition (null velocity at the liquid-solid interface), the vertical shear stress, $\tau$, in the fluid can be linearly approximated (Fig\ref{fgr:physics}.d) by:
\begin{equation}
 \tau = \mu \frac{\partial v}{\partial y} \sim \mu \frac{V_{wavefront}}{h},
\end{equation}
where $y$ is the distance perpendicular to the surface, $v$ is the fluid velocity in the direction parallel to the surface at height $y$, and $\mu$ is the dynamic viscosity of the medium (10$^{-3}$ m$^2$.s$^{-1}$ for PBS at 20$^{\circ}$C). In this formula, $V_{wavefront}$ and $h$ denote the surface velocity and the height of the thin film, respectively, estimated from a side-view image sequence (h $\sim$ 100 $\mu$m). Under SAW excitation, the "forward" wavefront velocity was about 100 mm/s, resulting in an estimated vertical shear stress of 1 Pa. When SAW excitation was turned off, the "backward" wavefront velocity was much lower ($\sim$ 10 mm/s), resulting in an estimated shear stress of 0.1 Pa. For comparison, values of shear stresses reported in the literature to detach cells from treated and untreated surfaces with flow-based systems typically lie between 0.01 and 10 Pa.\cite{Usami1993,Garcia1997,Didar2010,Hartmann2013a}

In addition to viscous stresses, pulsatile flow also induces so-called unsteady forces (Added mass). The relative magnitudes of viscous and unsteady forces can be quantified by the dimensionless Womersley number Wo = $\frac{\omega\rho h^2}{\mu}$. In our experiments, Wo was typically $\sim$ 0.5, meaning that unsteady forces likely also played an important role in cell detachment. 

\subsection{Cell detachment with single cell line (HEK 293)}
We first investigated the dynamics of cell detachment using our proposed method in experiments with a single cell line (HEK 293). The experiments sought to compare detachment behavior with substrates prepared with the two protocols described earlier: (1) cells adhered directly to the LiNbO$_3$ substrates after a 24 h incubation and (2) LiNbO$_3$ substrates incubated in a solution of re-suspended cells (incubation period of 60 min in this case). In the experiments, $\sim$ 300 cells were typically adhered initially to the surface in the field of view of the camera.

\begin{figure}[!h]
  \centering
  \includegraphics[height=13cm]{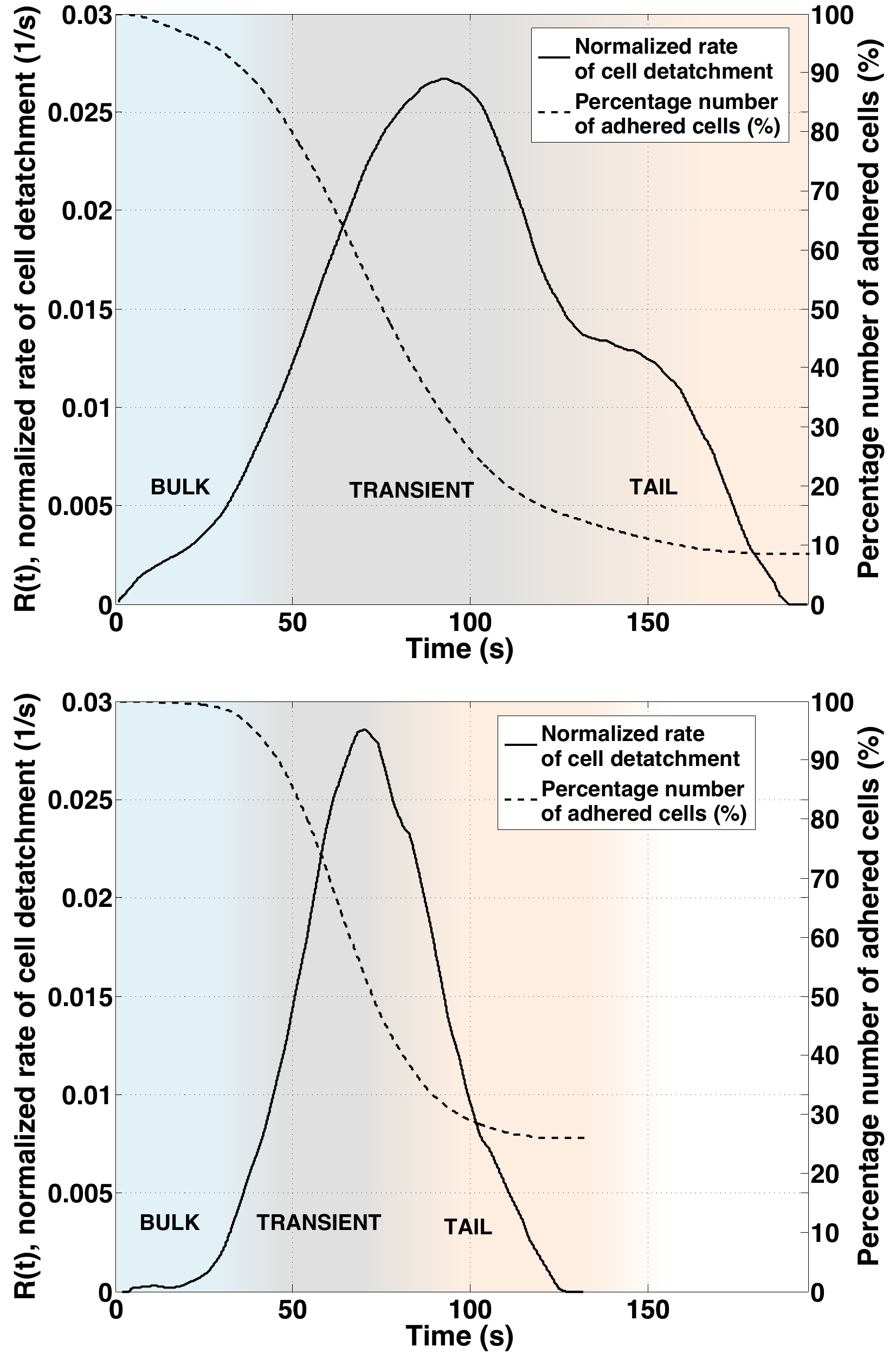}
  \caption{Normalized detachment rate and percentage of adhered cells over time under cyclic SAW actuation; TOP: result from a typical experiment with HEK 293 cells adhered directly to a LiNbO$_3$ substrate after a 24 h incubation; BOTTOM: result from a typical experiment with HEK 293 cells adhered to the substrate after incubation in a solution of re-suspended cells for 60 min.}
  \label{fgr:removal}
\end{figure}

Because the dynamics of cell detachment from a surface will vary greatly depending on particular conditions, experimental results should be compared in terms of normalized parameters. At a constant rate of detachment, the number of adhered cells will decrease by the same fraction over a time interval of fixed length at any point during the detachment process. As a result, the number of cells adhered to the surface as a function of time, $N(t)$, will follow a decaying exponential profile, $N(t) = N_0 e ^{-\sigma t}$, where $N_0$ is the initial number of attached cells and $\sigma$ is a detachment rate parameter. Cell detachment dynamics are characterized here in terms of their normalized rate of detachment from the surface, $R(t)$ :
\begin{equation}
 R(t) = - \frac{1}{N(t)} \frac{d N(t)}{dt}.
\end{equation}
The calculation of $R(t)$ is therefore equivalent to estimating the instantaneous value of the detachment rate parameter, $\sigma$, at time, t.

Fig.\ref{fgr:removal} shows measurements of the fraction of adhered cells over time and the calculated normalized detachment rates under cyclic SAW actuation for experiments with HEK 293 cells prepared with the two adhesion protocols (top: direct to LiNbO$_3$; bottom: incubation in re-suspended cells). The three fluid dynamics regimes (bulk, transient and tail) are highlighted by different background colors. Transitions between successive regimes were determined by observation in the image sequence where the top of the wavefront in the transient zone could be clearly seen. The uncertainty on estimations of the transition times was on the order of $\pm$5 s. The purpose of these experiments was to explore the relative differences in detachment kinetics between the three fluidic regimes and to extract "order of magnitude" information if possible about the dynamics. Indeed, a much broader range of experimental conditions would be required to form any kind of quantitative conclusion specific to these particular cell lines.

The results in Fig.\ref{fgr:removal} clearly show that the cell detachment rate is highest in the transient zone. The two graphs show similar maximum rates of detachment, indicating that both cell adhesion methods can yield similar adhesion strengths to the surface (assuming detachment rate is an indication of adhesion strength). The inverse of the maximum normalized detachment rate, $1/max(R(t))$, can be considered as the "characteristic time" of the system under maximum efficiency, that is to say, the time taken for the number of adhered cells to fall to $(1/e)N_0$ at the maximum rate of detachment. From Fig.\ref{fgr:removal}, this value is $\sim$ 35 s in both cases. The effect of the duration of the transient regime is also interesting to consider: in Fig5-top, the transient regime is longer ($\sim$ 63 s) resulting in a higher detachment efficacy ($>$ 10\% residual adhered cells), whereas in Fig.\ref{fgr:removal}-bottom due to a faster spreading of the droplet, the transient regime is shorter ($\sim$ 46 s) resulting in a lower detachment efficacy ($\sim$ 25\% residual adhered cells).  In both cases, the majority of cells is detached in minutes, which compares very favorably with results from others methods\cite{Hartmann2013a,Lu2004}.

\subsection{Cell detachment with dual cel lines (sorting)}
\begin{figure}[!h]
\centering
  \includegraphics[height=14cm]{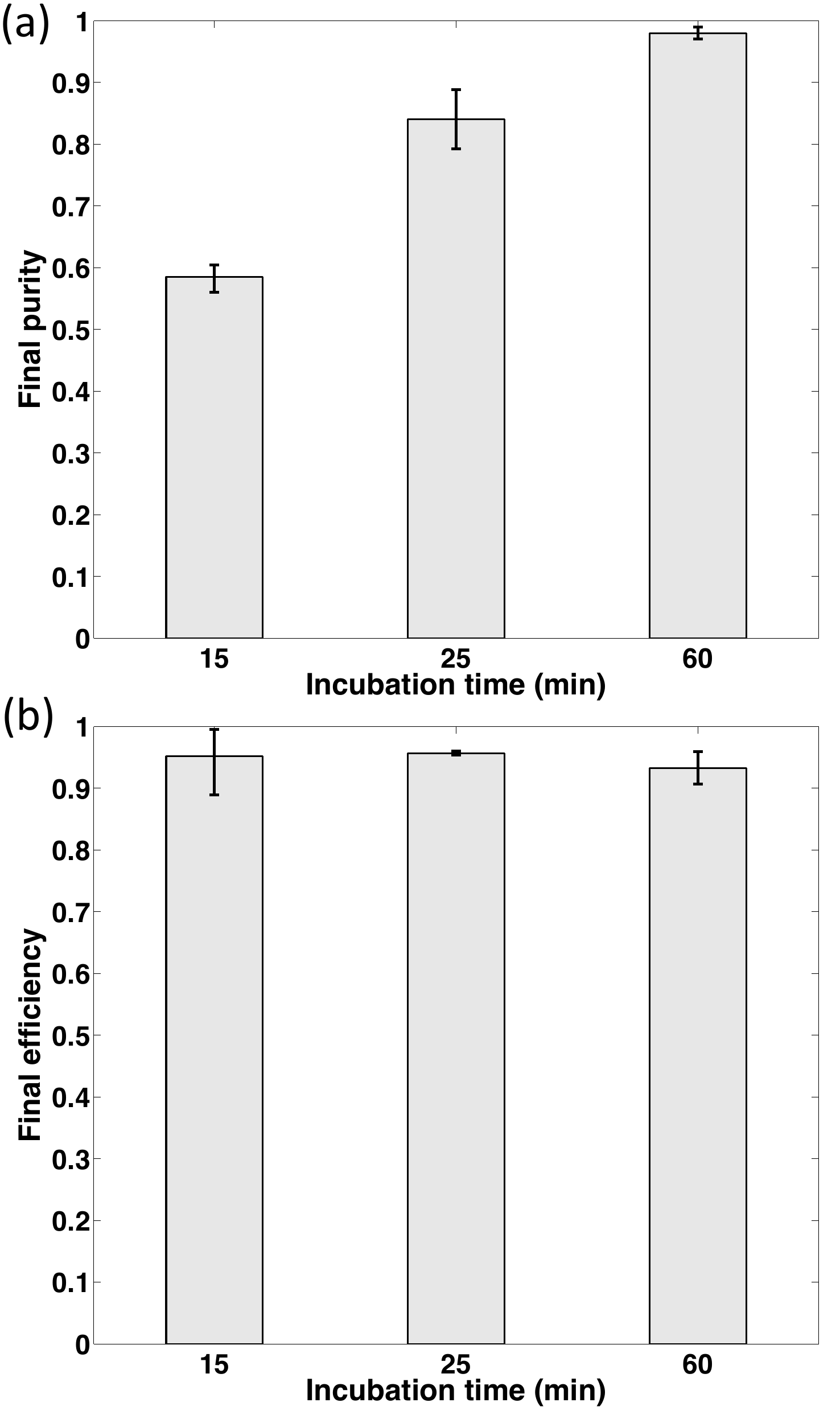}
  \caption{Cell sorting purity (a) and efficiency (b) for LiNbO$_3$ substrates prepared by incubation in a dual cell-line re-suspended solutions (A7r5 and HEK 293) for a range of incubation periods: 15, 25 and 60 min.}
  \label{fgr:sorting}
\end{figure}

We next investigated the selective detachment (sorting) of cell types based on adhesion contrast. LiNbO$_3$ substrates were prepared by incubation in a dual cell- solution of re-suspended cells line (A7r5 and HEK293) for a range of incubation periods: 15, 25 and 60 min. Cell sorting performance was characterized by 2 parameters, "purity" and "efficiency", calculated once the system had reached equilibrium, i.e. when the number of adhered cells no longer changed ($\sim$ 2 min typically):
\begin{equation}
 \mbox{Purity} = \frac{\%\mbox{ HEK 293}_{detached}}{\%\mbox{ HEK 293}_{detached}+\%\mbox{ A7r5}_{detached}}.
\end{equation} 
\begin{equation}
  \mbox{Efficiency} = \%\mbox{ HEK 293}_{detached}.
\end{equation}
Fig.\ref{fgr:sorting}a shows calculations of cell sorting purity for the three different incubation times. Interestingly, results indicate that purity increases with incubation time. We speculate that this behavior arises because: (1) cells require a certain time to achieve complete adhesion and thereby maximizing adhesion contrast between cell types and (2) it is possible that excretion of extracellular matrix proteins by A7r5 cells negatively modulates HEK 293 adhesion. Indeed, adhesion modulation by competing species has been observed by other groups, such as the improvement of cancer cell (MCF7) adhesion in the presence of human breast epithelial cells (MCF10A)\cite{Kwon2007}. In all cases, sorting efficiency was greater than 90 \% (Fig.\ref{fgr:sorting}b).

Once detached, cells remained re-suspended in the droplets. Short-term viability assays were performed after experiments both with SAW excitation and without SAW as a negative control. Results indicated that SAW excitation only slightly affected viability (apoptosis rate below 5 \%).

\section{Conclusions}
In this paper, we purpose a method to selectively and viably detach cells from a solid substrate in fluid droplets using modulated surface acoustic waves (SAW). Experiments were designed to study the effects of different fluid dynamics regimes by using a fixed imaging field of view with respect to droplet expansion under SAW actuation. Results show that cell detachment rate is highest in the middle regime, termed the "transient regime", where viscous shear stresses are estimated to be of the order of 1 Pa.

Under the chosen SAW modulation protocol, HEK 293 and A7R5 cells adhered to bare LiNbO$_3$ surfaces were successfully sorted based on adhesion contrast. Results show that cells were detached in the order of minutes and the contrast in adhesion strength varies with incubation time. This method could be generalized to other cell lines exhibiting either intrinsic or controlled (via surface bio-functionalization) adhesion contrasts. Importantly, cell adhesion strength is highly dependent on surface preparation specifics and adhesion modulation between competing species. Therefore, SAW excitation parameters required to viably detach and sort cell types under different experimental conditions can be expected to vary according to the characteristics of cell lines, surface preparation, and adhesion modulation between competing cell species. Similarly, sorting purity and efficiency are expected to be highly dependent on particular experimental conditions. 

Interestingly, modulated SAW could be combined with EWOD to detach strongly adhered cells and enhanced EWOD cell manipulation and sorting performance. Further investigations with a view to optimizing unsteady forces would also be of interest, for example with bi-lateral SAW excitation.

\section{Acknowledgements}
This work was supported by grants from the Natural Sciences and Engineering Research Council of Canada (NSERC), Cr\'eneau Biotech Santé du Projet ACCORD du Minist\`ere des Finances et de l'\'Economie (MFEQ) du Qu\'ebec (Canada), Agence Nationale pour la Recherche ANR-12-BS09-021-02 et 01 (France), Direction G\'en\'erale de l'Armement (France) and R\'egion Nord-Pas-de-Calais (France). The authors would like to thank the Institut Interdisciplinaire d'Innovation Technologique (3IT) at the Universit\'e de Sherbrooke (Canada) for technical support.

\footnotesize{
\bibliography{biblio} 
\bibliographystyle{rsc} 
}

\end{document}